%% file: sb.tex
\documentclass[12pt]{iopart}
\usepackage{iopams}  
\usepackage{epsfig,rotating}
\input mydefcp.tex

\newcommand{\thetaot}{\ensuremath{\theta_{13}}\,}

\begin{document}
\title{\bf Physics Potential of the SPL Super Beam }

\author{Mauro Mezzetto }

\address{ Istituto Nazionale Fisica Nucleare, Sezione di Padova, Italy.
\vskip0.5cm Invited talk at the Nufact02 Workshop, Imperial College of Science, Technology and Medicine, London,
July 2002.}

\begin{abstract} 
Performances of a neutrino beam generated by the CERN SPL proton driver
are computed considering a 440 kton water \v{C}erenkov detector at 130 km from
the target.
$\theta_{13}$ sensitivity down to $1.2^\circ$ and a $\delta$ sensitivity comparable to
a Neutrino Factory, for $\theta_{13} \geq 3^\circ$, are within the reach
of such a project.
\end{abstract}

\maketitle

\section{Introduction}
The planned Super Proton Linac (SPL) 
 is a 2.2 GeV proton beam of 4 MW power \cite{SPL}
working with a repetition rate
of 75 Hz delivering $1.5\cdot10^{14}$ protons per pulse ($10^{23}$ protons
on target (pot) in a $10^{7}$ s conventional year).
It could be the first stage
of a CERN based Neutrino Factory or of a Beta Beam.

Studies of the capabilities of a neutrino
beam generated by SPL have been already published in \cite{myself}, \cite{yr},
in this paper the fluxes and the overall physical performances 
will be reviewed in the light of the new design of
the beam optics, specially optimized for the SuperBeam needs \cite{Simone}.
They are computed for a gigantic
water \v{C}erenkov detector, as the proposed UNO detector \cite{UNO} 
(440 kton fiducial)
located in the Modane laboratory
under the Frejus tunnel at a baseline of 130 km from CERN.

The SPL SuperBeam capabilities 
would constitute a natural follow-up of the JHF phase I experiment \cite{JHF},
with excellent sensitivity on \thetaot (section~3) and good
sensitivity on the CP phase $\delta$ (section~4).
Furthermore the SPL SuperBeam could be used to complement
the results of a Neutrino Factory experiment, helping in resolving
the ambiguities, as discussed in \cite{Mena}; or could be combined
with a Beta Beam, as discussed in \cite{mybeta}.

Signal efficiency and backgrounds have already been discussed in \cite{myself},
they are computed by using the NUANCE neutrino event generator
\cite{NUANCE} and reconstructing the events with standard SuperKamiokande
algorithms, with the addition of improved $\pi^\circ$ rejection algorithms.
They can be summarized as signal efficiency $\epsilon\simeq 70\%$ and
$\pi^\circ$ and $\mu/e$ 
background rejection, normalized to the non oscillated \numu charged current
interactions, 
 $f_B^{\pi^\circ}=4.2\cdot10^{-4}$,
 $f_B^{\mu/e}=3\cdot10^{-3}$.

\section{Fluxes}
Details of the new beam optics can be found in \cite{Simone}. The use
of an horn and a reflector increases by $\sim 40\%$ the overall \numu
flux with respect to the former single horn optics,
 slightly increases the \nue contamination, while the \nubarmu and
\nubare contaminations are reduced by $\sim 30\%$.

\begin{figure}[th]
  \epsfig{file=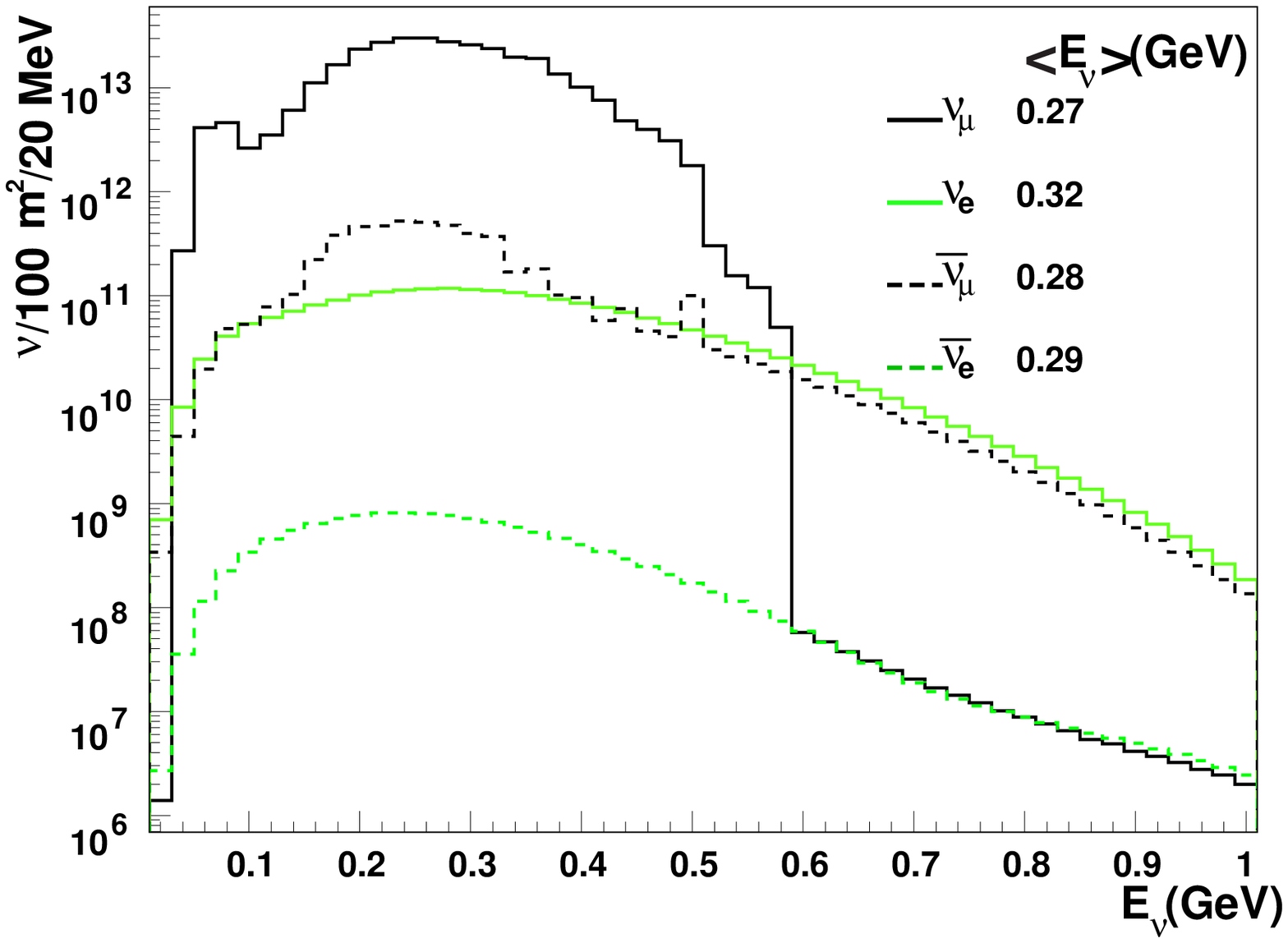,width=0.51\textwidth}\hfill
  \epsfig{file=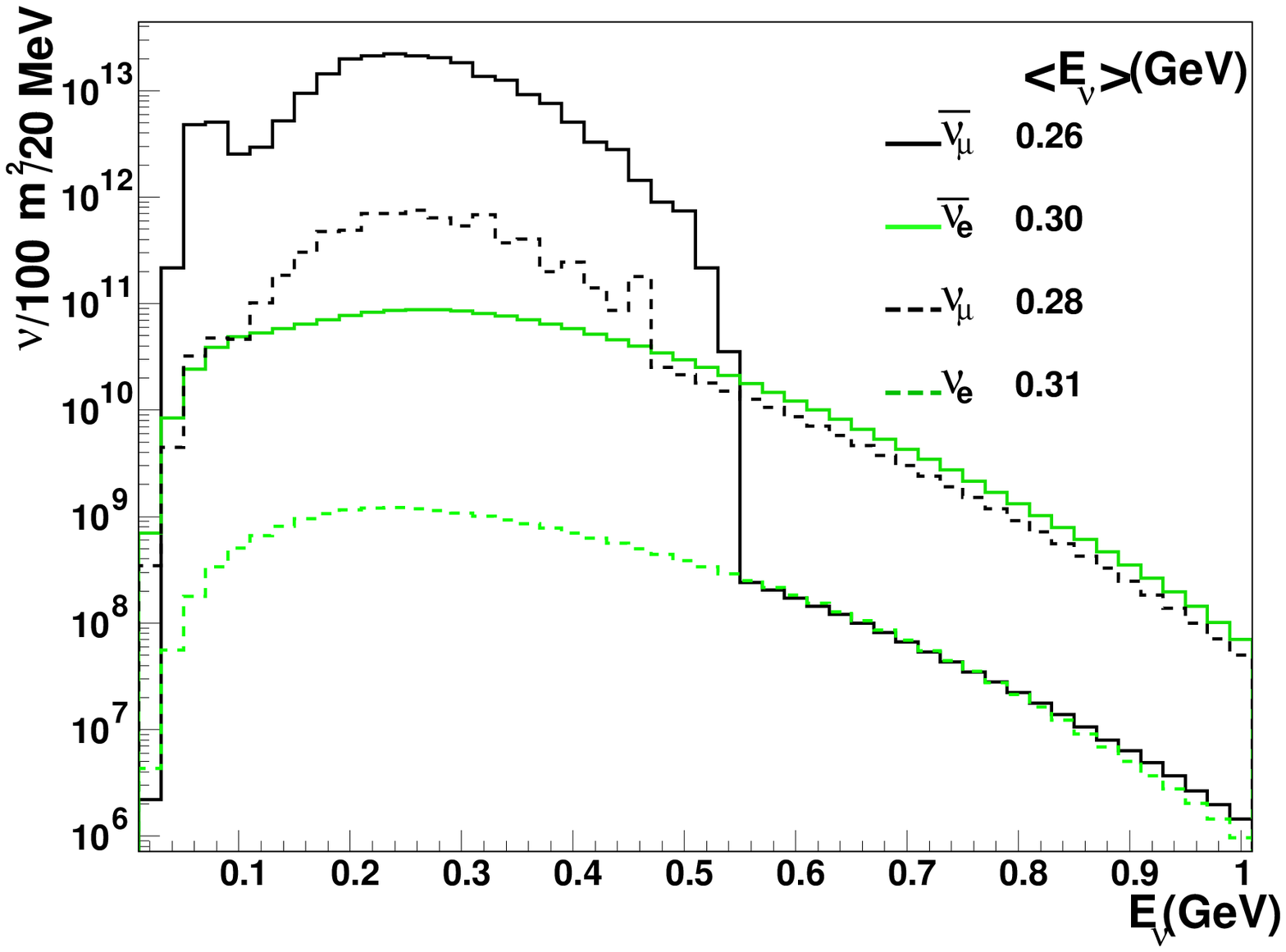,width=0.51\textwidth}
\caption{ Neutrino spectra for the $\pi^+$ (left) and the $\pi^-$
(right) focussed beam for a decay tunnel length of 60 m.}
\label{fig:spectra}
\end{figure}

The length of the decay tunnel has been re-optimized having in mind
CP searches more than \thetaot. Table \ref{tab:fluxes} reports details of the
beam properties as function of the length of the decay tunnel,
including the sensitivity on \thetaot for a 2200 kton/year exposure.
In spite of the fact that the \thetaot sensitivity is maximum for the
lowest length (20 m), a 60 m decay length is preferred because
of the lower \nubarmu contamination, that results in a better CP
sensitivity.
\Table{\label{tab:fluxes}Neutrino fluxes and contamination for different
values of the decay tunnel length. The last line refers of the
single horn optics of ref.~\cite{myself}.  }
\begin{tabular}{cccccccc}
\br
       & \multicolumn{3}{c}{$\pi^+ focus$} & \multicolumn{3}{c}{$\pi^- focus$} &
$\pi^+ focus$ \\
\br
Length & \numu & \nue & \nubarmu & \nubarmu & \nubare & \numu & \thetaot \\
(m) & ($\nu/m^2/yr$) & (\%) & (\%) & ($\nu/m^2/yr$) & (\%) & (\%) & (90\%CL) \\
    & (@50 km) &    &            & (@50 km) &    &  &(2200 kton/yr)   \\
\hline
20 & 2.43 $\cdot 10^{+12}$& 0.38 & 1.71 &1.73$\cdot 10^{+12}$ & 0.41 & 3.9 & 1.20\\
60 & 3.23 $\cdot 10^{+12}$& 0.67 & 1.50 & 2.25$\cdot 10^{+12}$ & 0.70 & 3.3 & 1.25\\
100 & 3.35 $\cdot 10^{+12}$& 0.76 & 1.62 & 2.33$\cdot 10^{+12}$ & 0.79 & 3.3 & 1.30\\
\hline
20 (old) & 1.71$\cdot 10^{+12}$ &0.36& 2.4&1.12$\cdot 10^{+12}$ & 0.38 & 5.6 & 1.47 \\
\br
\end{tabular}
\endTable
The neutrino spectra for the $\pi^+$ and $\pi^-$ focussed beams are
displayed in Fig.~\ref{fig:spectra}.
\section{Sensitivity on \thetaot}
The \thetaot sensitivity is computed assuming $\delta=0$, solar SMA solution,
$\delta m^2_{23}=2.5\cdot 10^{-3}$ eV$^2$,  $\theta_{23}=45^\circ$ and 5 years
of data taking.
These are the standard benchmark assumptions used by similar projects
\cite{JHF}, \cite{othertheta}.

\begin{figure}[tbh]
\begin{minipage}{0.68\textwidth}
\epsfig{file=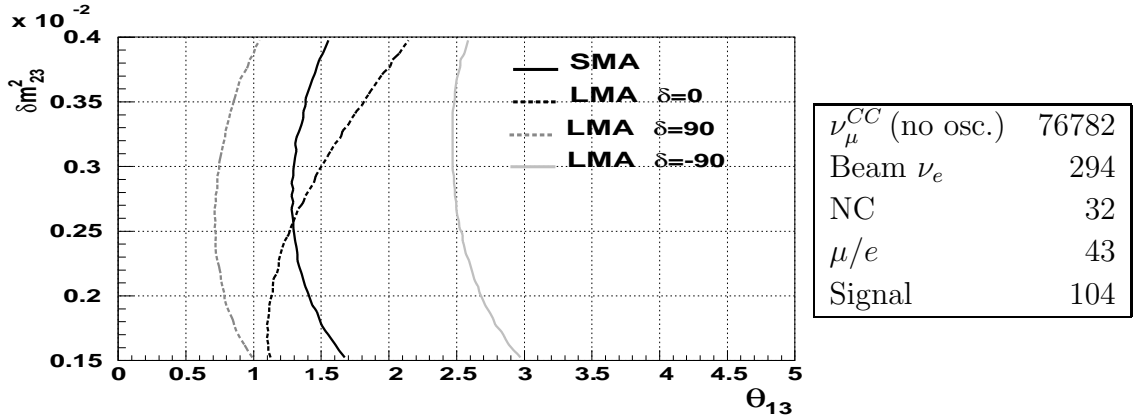,width=\textwidth}
\end{minipage}
\begin{minipage}{0.30\textwidth}
\begin{tabular}{|lr|}
\hline
\numucc (no osc.) & 76782 \\
Beam \nue & 294 \\
NC & 32 \\
$\mu/e$ & 43 \\
Signal & 104  \\
\hline
\end{tabular}
\end{minipage}
\caption{Left: \thetaot sensitivity (90\%CL) computed for
a 2200 kton exposure, under the solar SMA solution or under LMA and
different  $\delta$ values.
Right: Number of events for the same exposure, SMA solution, in case of
$\thetaot=2^\circ$.}
\label{fig:thetaot}
\end{figure}

Fig.~\ref{fig:thetaot} shows the \thetaot sensitivity (90\% CL) in case of no 
signal and summarizes the event rate
computed for \thetaot=$2^\circ$.
 The experiment would have sensitivity down to $\thetaot=1.2^\circ$
($\sin^2{2\thetaot}=1.75\cdot10^{-3}$) 
%
\section{CP sensitivity}
CP sensitivity is computed assuming a 2 year run with the $\pi^+$ focussed
beam and 8 years with the $\pi^-$ focussed beam. This sharing is motivated by the
unfavorable cross section ratio $\nubare/\nue \sim 1/6$ at 300 MeV.

        A 10\% error on the solar $\delta m^2$ and $\sin^2{2\theta}$,
as expected from the KamLAND experiment~\cite{Kamland} and
  a 2\% error on the atmospheric $\delta m^2$ and $\sin^2{2\theta}$,
as expected from the JHF neutrino experiment~\cite{JHF} are taken into account.
Correlations between \thetaot and $\delta$ are fully accounted for,
while the sign $\delta m^2_{13}$ and the
$\theta_{23}/(\pi/2-\theta_{23})$ ambiguities are not considered.
A systematic error of 2\% is accounted for the signal efficiency and
background normalization, as discussed in \cite{myself}.

\begin{figure}[tb]
{\epsfig{file=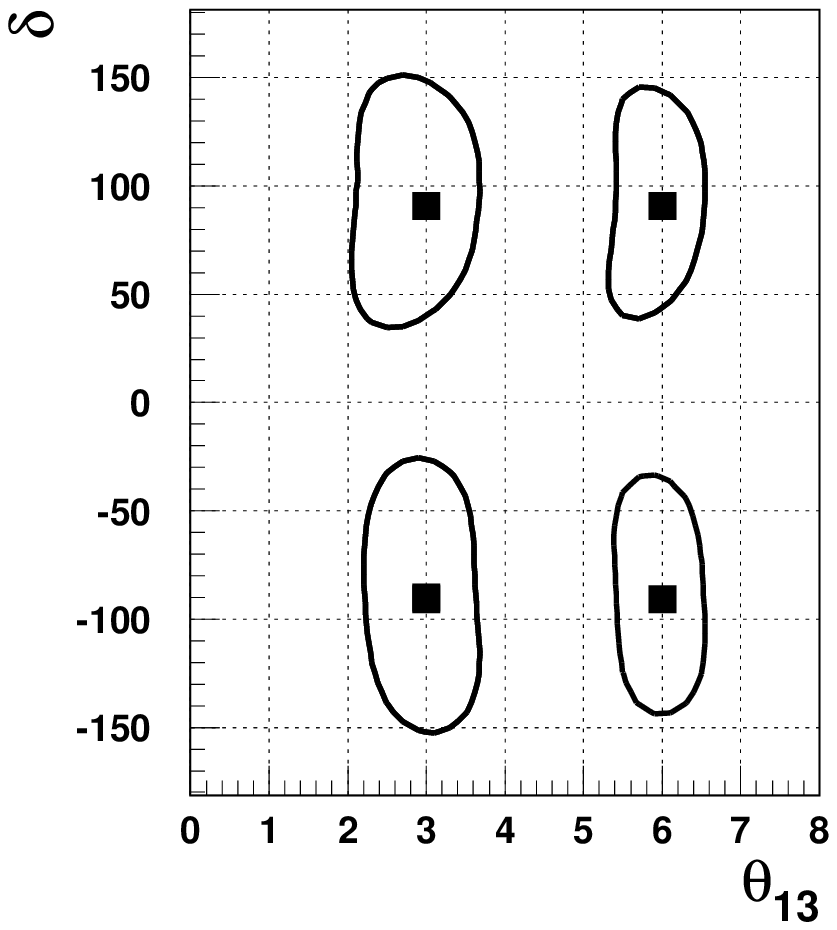,width=0.44\textwidth} }
\hspace*{-0.5cm}
{\epsfig{file=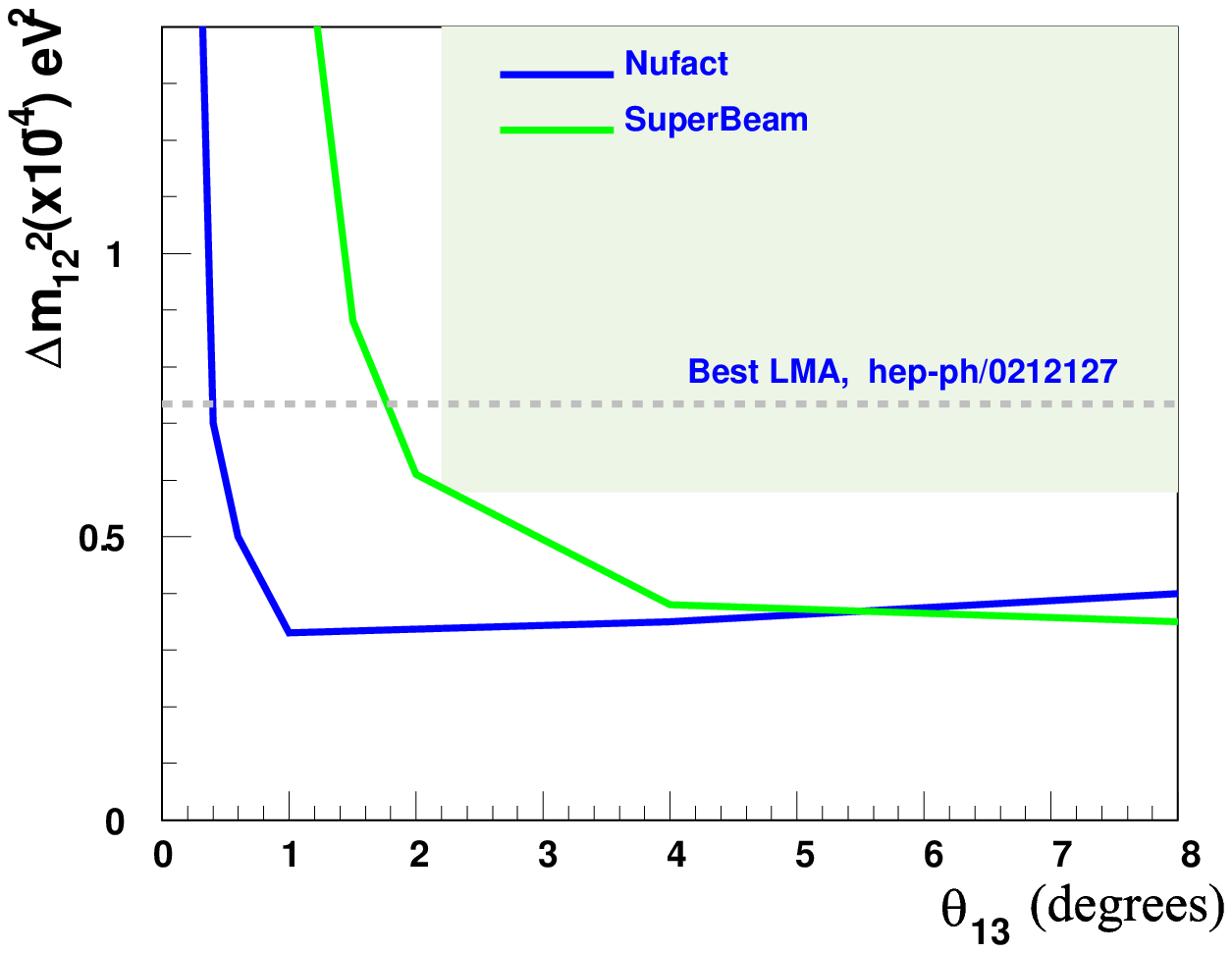,width=0.62\textwidth} }
\caption{Left: $\thetaot - \delta$ fits (99\% CL)  computed for
$\delta m^2_{12}=10^{-4}$ eV$^2$, $\sin^2{2 \theta_{12}}=0.8$.
The squares indicate the starting points.
Right: CP sensitivity of  the SPL-SuperBeam, see text,
 compared with a 50 GeV
Neutrino Factory 
producing       $2\cdot10^{20} \mu$ decays/straight section/year, 
and two 40 kton detectors at 3000 and 7000 km \cite{golden};
the shaded region corresponds to the allowed LMA solution and the
\thetaot sensitivity of JHF.}
\label{fig:CP}
\end{figure}

Solutions for different values of $\delta$ and \thetaot, 
Fig.\ref{fig:CP}-left, show
very small correlations between the two parameters.

Since the sensitivity to CP violation heavily depends on the true 
value of \dmot and \thetaot, we prefer to express the CP sensitivity for a
fixed value of $\delta$ in the \dmot, \thetaot parameter space.
The CP sensitivity to separate
$\delta=90^\circ$ from $\delta=0^\circ$ at the 99\%CL as a function
of \dmot and \thetaot, following the convention of~\cite{golden},
is plotted in Fig.~\ref{fig:CP}-right.
 
It is fair to say that
SPL SuperBeam CP sensitivity approaches  the Neutrino Factory sensitivity 
in the parameter space that will be explored by the JHF experiment:
$\thetaot \geq 2.3^\circ$.

\newpage

\end{document}

%% file: mydefcp.tex
\def\dm{\ensuremath{\Delta m}}
\def\dm2{\ensuremath{\Delta m^{2}\ }}
\def\sen2th{\ensuremath{ \sin^{2}(2\theta)\ }}
\def\(({\left(}
\def\)){\right)}

\def\nue{\ensuremath{\nu_{e}\ }}
\def\nubare{\ensuremath{\overline{\nu}_{e}\ }}

\def\numu{\ensuremath{\nu_{\mu}\ }}
\def\nubarmu{\ensuremath{\overline{\nu}_{\mu}\ }}

\newcommand{\numucc}{\ensuremath{\nu_\mu^{CC}\,}}

\def\mc2{\multicolumn{2}{c|}}
\newcommand{\dmot}{\ensuremath{\delta m^2_{12}\,}}